\documentclass[numsec,webpdf,modern,medium,namedate]{oup-authoring-template}

\makeatletter
\renewcommand*\l@subsubsection[2]{{}{}{}} 
\makeatother
\usepackage{subcaption}
\usepackage{booktabs}

\newlength{\dhatheight}

\newcounter{probnum}
\allowdisplaybreaks

\definecolor{tabblue}{rgb}{.870588,.905882,.94902}
\definecolor{gray}{rgb}{0.7,0.7,0.7}
\definecolor{black}{rgb}{0,0,0}
\definecolor{white}{rgb}{1,1,1}
\definecolor{blue}{rgb}{0.0,0.0,1}

\definecolor{green}{rgb}{0,0.5,0}

\definecolor{yellow}{rgb}{1,0.549,0}

\definecolor{red}{rgb}{0.6,0.0,0.0}

\definecolor{darkred}{rgb}{0.9,0.4,0}

\definecolor{purple}{rgb}{0.58,0,0.827}

\definecolor{backgcode}{rgb}{0.97,0.97,0.8}
\definecolor{Brown}{cmyk}{0,0.81,1,0.60}
\definecolor{OliveGreen}{cmyk}{0.64,0,0.95,0.40}
\definecolor{CadetBlue}{cmyk}{0.62,0.57,0.23,0}

\newcommand{\qu}[1]{``{#1}''}
\newcommand{\bv}[1]{\boldsymbol{#1}}

\newcommand{\tausq}{\tau^2}

\newcommand{\bSigma}{\bv{\Sigma}}

\newcommand{\B}{\bv{B}}

\newcommand{\X}{\bv{X}}

\newcommand{\I}{\bv{I}}

\newcommand{\Y}{\bv{Y}}

\newcommand{\x}{\bv{x}}

\newcommand{\w}{\bv{w}}

\newcommand{\y}{\bv{y}}

\renewcommand{\b}{\bv{b}}

\newcommand{\bbeta}{\bv{\beta}}

\newcommand{\twovec}[2]{\bracks{\begin{array}{c} #1 \\ #2 \end{array}}}

\newcommand{\reals}{\mathbb{R}}

\newcommand{\beqn}{\begin{eqnarray*}}
\newcommand{\eeqn}{\end{eqnarray*}}
\newcommand{\bneqn}{\begin{eqnarray}}
\newcommand{\eneqn}{\end{eqnarray}}
\newcommand{\benum}{\begin{enumerate}}
\newcommand{\eenum}{\end{enumerate}}

\newcommand{\parens}[1]{\left(#1\right)}

\newcommand{\tothepow}[2]{\parens{#1}^{#2}}
\newcommand{\prob}[1]{\mathbb{P}\parens{#1}}

\newcommand{\cprob}[2]{\prob{#1~|~#2}}

\newcommand{\bracks}[1]{\left[#1\right]}
\newcommand{\braces}[1]{\left\{#1\right\}}

\newcommand{\inverse}[1]{\parens{#1}^{-1}}

\newcommand{\varnostr}[1]{\mathbb{V}\text{ar}[#1]}

\renewcommand{\exp}[1]{\mathrm{exp}\parens{#1}}

\renewcommand{\sin}[1]{\text{sin}\parens{#1}}

\newcommand{\oneover}[1]{\frac{1}{#1}}

\newcommand{\multnormnot}[3]{\mathcal{N}_{#1}\parens{#2,\,#3}}

\newcommand{\normnot}[2]{\mathcal{N}\parens{#1,\,#2}}

\newcommand{\uniform}[2]{\mathrm{U}\parens{#1,\,#2}}

\newcommand{\invgammanot}[2]{\text{Inv}\mathcal{G}\parens{#1,\,#2}}

\newcommand{\errorrv}{\mathcal{E}}

\newcommand{\bBconcord}{\B_{\mathcal{C}}}

\newcommand{\bbconcord}{\b_{\mathcal{C}}}

\newcommand{\bSigmaconcord}{\bSigma_{\mathcal{C}}}

\newcommand{\Xconcord}{\X_{\mathcal{C}}}
\newcommand{\yconcord}{\y_{\mathcal{C}}}
\newcommand{\Xdiscord}{\X_{\mathcal{D}}}
\newcommand{\ydiscord}{\y_{\mathcal{D}}}

\usepackage{amsmath, amssymb}
\usepackage[fontsize=12pt]{fontsize}

\newcommand{\ourtitle}{Improved Conditional Logistic Regression using Information in Concordant Pairs with Software}
\newcommand{\ourpackage}{\texttt{bclogit}}

\begin{document}

\journaltitle{Journal}
\DOI{DOI added during production}
\copyrightyear{2026}
\pubyear{2026}
\vol{XX}
\issue{x}
\access{Published: Date added during production}
\appnotes{Paper}

\firstpage{1}

\title[Bayesian Conditional Logistic Regression]{\ourtitle}

\author[1]{Jacob Tennenbaum\ORCID{0009-0006-1408-5035}}
\author[1,$\ast$]{Adam Kapelner\ORCID{0000-0001-5985-6792}}

\address[1]{\orgdiv{Mathematics}, \orgname{Queens College, CUNY}, \orgaddress{\street{65-30 Kissena Blvd}, \postcode{11367}, \state{Flushing, NY}, \country{USA}}}

\corresp[$\ast$]{Corresponding author: \href{mailto:kapelner@qc.cuny.edu}{kapelner@qc.cuny.edu}}

\received{Date}{0}{Year}
\revised{Date}{0}{Year}
\accepted{Date}{0}{Year}

\abstract{We develop an improvement to conditional logistic regression (CLR) in the setting where the parameter of interest is the additive effect of binary treatment effect on log-odds of the positive level in the binary response. Our improvement is simply to use information learned above the nuisance control covariates found in the concordant response pairs' observations (which is usually discarded) to create an informative prior on their coefficients. This prior is then used in the CLR which is run on the discordant pairs. Our power improvements over CLR are most notable in small sample sizes and in nonlinear log-odds-of-positive-response models. Our methods are released in an optimized \texttt{R} package called \ourpackage.}






\maketitle\onecolumn


\section{Introduction}\label{sec:introduction}

Consider the setting of $2n$ observations each with $p$ covariate measurements denoted as the row vectors $\x_1, \ldots, \x_{2n} \in \reals^p$ and stacked rowwise into the matrix $\X \in \reals^{n \times p}$. Each observation is assigned (or observed to have) a treatment denoted as $w_i$. Herein we will be considering the case of a binary treatment $w_1, \ldots, w_{2n} \in \braces{0,1}$ stacked into a column vector $\w$ where $w_i = 0$ will be synonymously termed \qu{control} and $w_i = 1$ will be synonymously termed \qu{treatment}. Further, the measured response for each observation is binary, i.e., $y_1, \ldots, y_{2n} \in \braces{0,1}$ stacked into a column vector $\y$ where $y_i = 1$ will be synonymously termed a \qu{positive response}. We will denote $Y_1, \ldots, Y_{2n}$ stacked into a column vector $\Y$ as the Bernoulli random variables responsible for realizing the responses. We then assume a model for the log-odds of the positive response to be linear in the conditioned covariate measurements and the treatment effect, 
\beqn
p_i := \cprob{Y_i}{\beta_w, \beta_0, \bbeta, w_i, \x_i} = (1 + e^{-(\beta_0 + \x_i \bbeta + \beta_w w_i)})^{-1}.
\eeqn
%
\noindent Assuming all $2n$ responses are independent, the likelihood function can be written as
\bneqn\label{eq:logistic_regression_likelihood}
\cprob{\Y}{\beta_w, \beta_0, \bbeta, \w, \X} = \prod_{i=1}^{2n} p_i^{y_i} \tothepow{1 - p_i}{1 - y_i}
\eneqn
\noindent where $\beta_0 \in \reals, \bbeta \in \reals^p$ denote the $p+1$ nuisance parameters and $\beta_w \in \reals$ represents the parameter of interest: the additive effect on log-odds of positive response of the treatment. Estimation of all $p+2$ parameters is via the iterative weighted least squares procedure in standard logistic regression \citep[LR,][]{walker1967estimation}.

In both observational and experimental research, data often emerge in inherent pairings where the two units are so closely linked that they share a vast majority of their underlying characteristics. This can occur when data are measured on human siblings (especially monozygotic twins), before/after (pre-test / post-test) designs where each individual's binary response is measured at two different times, contra-lateral studies where response measurements are made on an individual's left/right eyes (or left/right limbs, etc), social science studies where measurements are made on both the husband and wife or on each roommate in a two-person living arrangement, etc. 

Let our data now be observed as $n$ pairs. Without loss of generality, let the pairs (sometimes referred to by the general term \qu{strata}) be organized as observation indices $<1,2>, <3,4>, \ldots, <2n-1,2n>$ where the first observation in the pair receives $w=1$ (i.e., the odd numbered indices) and the second observation in the pair receives $w=0$ (i.e., the even numbered indices). In this setting, the likelihood function in~\eqref{eq:logistic_regression_likelihood} is incorrect as it omits the many more covariate measurements shared among the intrapair subjects (beyond the $p$ observed covariates $\x_i$ measured on each subject). These unobserved characteristics induce an intrapair dependence. 

To correct the likelihood function, there are many strategies. One could add different correlations for each pair $\rho_1, \ldots, \rho_n$ but then the number of parameters expands at the same rate as the number of observations. This is called the \qu{incidental parameter problem} and results in biased and inconsistent estimation. To shrink the parameter space to not grow with $n$, one commonly assumes the simplified setting of one shared intrapair correlation $\rho$ for all $n$ pairs. Estimation then follows from the Generalized Estimating Equations (GEE) of \citet{liang1986longitudinal,zeger1986longitudinal}. Equivalently, the effect of the unobserved covariates can be distilled into a separate intercept for each pair $\alpha_1, \ldots, \alpha_n$. Again, the number of parameters expands at the same rate as the number of observations. To shrink the parameter space, one can assume that the intercepts are drawn randomly, e.g. from a normal distribution centered at 0 with unknown variance and this variance becomes the one nuisance parameter. Estimation then uses the generalized linear mixed model (GLMM) of \citet{stiratelli1984random, breslow1993approximate}. A fourth strategy, and our focus herein, is the Conditional Logistic Regression (CLR) model of \citet{prentice1978retrospective} explained below. 

Consider pairs of observations $<j,k>$ where the response of one observation is positive and the response of the other observation is negative, i.e. $y_j +y_k = 1$. The two possible likelihoods for such pairs can be shown through elementary algebra \citep[e.g., in][Section 11.2.6]{agresti2013categorical} to be
\bneqn\label{eq:discordant_pairs_clr}
\cprob{Y_j = 1, Y_k = 0}{Y_j + Y_k = 1, \beta_w, \bbeta} &=& \oneover{1 + e^{(\x_k - \x_j) \bbeta + \beta_w}}, \nonumber \\
\cprob{Y_j = 0, Y_k = 1}{Y_j + Y_k = 1, \beta_w, \bbeta} &=& \oneover{1 + e^{(\x_j - \x_k) \bbeta + \beta_w}} ~~~~~~
\eneqn
\noindent and are aggregated together by multiplying across all such pairs as pairs are independently realized. Such pairs above are called \qu{discordant} as their two responses are not equal (literally, their responses do not accord with one another). The likelihoods of the \qu{concordant} pairs (i.e., those where $y_j + y_k \in \braces{0,2}$) conditional on their sum $Y_j + Y_k$ can be shown to be 1. Thus, observations in the concordant pairs do not factor into the likelihood function and, thus, traditionally are considered ancillary adding no information to the inference of $\beta_w$. 

The conceptual origins of CLR can be traced to \citet{penrose1934relative} who used a conditional likelihood approach to isolate the effects of parental age on Down syndrome risk. By conditioning the analysis on the sibship, there was effective control for birth order and other familial confounders. This idea seemed to remain dormant until the 1970s, when it was independently reinvented and formalized. In the field of economics, \citet{mcfadden1973conditional} developed the conditional logit model to address discrete choice behavior, where the \qu{stratum} represented a decision-maker and the observations were the various alternatives available to them. In epidemiology, researchers match cases to controls on a one-to-one or one-to-many basis, they create numerous strata. To solve the incidental parameter problem we described above, \citet{prentice1978retrospective} adapted the partial likelihood principles of \citet{cox1972regression}'s proportional hazards model to retrospective data. By conditioning the likelihood on the total number of cases within each matched set, they found the stratum-specific intercepts $\alpha_i$ cancel, yielding the elegant expressions found in~\eqref{eq:discordant_pairs_clr}. The practical standardization of this method was achieved by \citet{breslow1980statistical}, whose definitive work established CLR as standard methodology. While primarily associated with case-control studies, the utility of CLR extends naturally into experimental design, particularly in within-subject experiments and randomized matched-pair trials (our setting). 

Our contribution is conceptually simple but powerful as it is costless. Instead of discarding the observations within the concordant pairs, we salvage the information in this data via a two stage estimation: (a) pre-modeling with the concordant data to estimate the nuisance parameters $\bbeta$ and (b) use these estimates to specify an empirical Bayes prior for a Bayesian conditional logistic regression (BCLR). This prior information on $\bbeta$ sharpens the estimation for the parameter of interest $\beta_w$. Section~\ref{sec:methodology} details our methodological contribution, Section~\ref{sec:simulations} demonstrates our methods' advantages over models common in standard practice and Section~\ref{sec:discussion} concludes, offering advice to practitioners and discusses further directions.

\section{Methodology}\label{sec:methodology}

\subsection{Setup}\label{subsec:methodology_setup}

Let $\mathcal{C}$ denote the set of concordant pairs of observations and let $\mathcal{D}$ denote the set of disconcordant pairs of observations so that $\mathcal{C} \cup \mathcal{D} = \braces{<1,2>, \ldots, <2n-1,2n>}$ and $|\mathcal{C}| + |\mathcal{D}| = n$. Denote $\bBconcord$ and $\bbconcord$ as an estimator and estimate of $\bbeta$ respectively. In a premodeling step, we estimate $\bbconcord$ and $\bSigmaconcord := \varnostr{\bBconcord}$ using data $\yconcord$, $\Xconcord$. Each pair in $\mathcal{C}$ is treated as a stratum, $\w$ is dropped (as its parameter is zero) and the nuisance intercept parameter is included.

To do this premodeling step, we consider fitting with three modeling choices: LR, GEE and GLMM. The first two of these choices have systemic problems. The LR ignores the dependence structure of $\mathcal{C}$. The GEE model is slightly incoherent as its population-level interpretation of $\bbeta$ will clash with the subject-level interpretation of $\bbeta$ in CLR. We further discuss our recommendation choice in the concluding Section~\ref{sec:discussion} after first seeing the simulations of Section~\ref{sec:simulations}.

Regardless of the modeling choice employed, we then use the premodel's estimates $\bbconcord$ and $\bSigmaconcord$ as hyperparameters in one of four informative prior distributions (see following subsections). BCLR is then defined as performing Bayesian estimation of $\bbeta$, $\beta_w$ using one of those priors, denoted $\pi$. We will partition the posterior's kernel as
\beqn
f(\beta_w, \bbeta\,|\,\y,\,\X) \propto p(\ydiscord\,|\,\beta_w, \bbeta, \Xdiscord)  \pi(\beta_w, \bbeta\,|\,\yconcord,\X)
\eeqn
\noindent where the discrete likelihood above is given in \eqref{eq:discordant_pairs_clr}. We then sample $f(\beta_w, \bbeta\,|\,\y,\,\X)$ via a Markov Chain Monte Carlo (MCMC). By considering only the MCMC samples of $\beta_w$, we have inference for our parameter of interest averaged over the uncertainty in the nuisance parameters $\bbeta$ \citep{gelfand1990sampling}. For confidence set construction for $\beta_w$, we will consider an equal-tailed $1 - \alpha = 95\%$ credible region denoted CR$_{\beta_w, 1 - \alpha}$, an idea first found in \citet{lindley1965inference}, which was noted to be a transformation-invariant decision rule by \citet{gelman2013bayesian}. For two-sided hypothesis testing of $H_0: \beta_w = \theta_0$, we consider a rejection defined as $\theta_0 \notin$ CR$_{\beta_w, 1 - \alpha}$.

The rest of this section is devoted to four different specifications of the prior $\pi$, which describe four iterative improvements to CLR. 


\subsection{The Naive Prior}\label{subsec:prior:naive}

We employ a multivariate normal (MVN) as recommended by \citet{GelmanHillVehtari2020} as they considered the MVN a good starting point for a weakly informative prior, as it provides some regularization without being \qu{stuck} in heavy tails. We use a flat prior for the treatment effect parameter and assume independence between it and the covariate effects,
\bneqn\label{eq:Normal_prior_without_intercept_flat_treatment}
\pi\parens{\beta_w, \bbeta\,|\,\yconcord,\X} = \multnormnot{p+1}{
    \twovec{\beta_w}{\bbeta};\,\twovec{0}{\bbconcord}
}{\begin{bmatrix}
        \tausq & 0\\
        0 & \bSigmaconcord
    \end{bmatrix}}
\eneqn
\noindent where $\tausq$ is chosen to be a large positive number during implementation.

\subsection{The Mixture of g's prior}\label{subsec:prior:g}

The prior in \eqref{eq:Normal_prior_without_intercept_flat_treatment} does not allow for calibrating the strength of $\bbeta$ in the concordant data on the posterior. To do so, we consider the mixture-of-$g$'s-prior  \citep{zellner1986, liang2008mixtures},
\bneqn\label{eq:g_prior}
\pi(\beta_w,\bbeta,\,g\,|\,\yconcord,\X) &=& 
\pi(\beta_w, \bbeta\,|\,g,\,\,\yconcord,\X)
\pi(g\,|\,\yconcord, \X) \nonumber\\
&=&
\multnormnot{p+1}{
\twovec{\beta_w}{\bbeta};\,\twovec{0}{\bbconcord}
}{\begin{bmatrix}
        \tausq & 0\\
        0 & g\bSigmaconcord
    \end{bmatrix}} \times  \nonumber\\
&& \invgammanot{g;\,\frac{1}{2}}{\frac{|\mathcal{D}|}{2}}. 
\eneqn
\noindent The choice of the inverse gamma hyperprior implies a marginal multivariate Cauchy distribution for $\bbeta$ (as originally found in \citealt{zellner_siow_1980}). Fitting $g$ using both the concordant data and the size of the discordant data can provide a more flat prior than the MVN prior of \eqref{eq:Normal_prior_without_intercept_flat_treatment}.

\subsection{The Probability-Matching Prior (PMP)}\label{subsec:prior:pmp}

As properly-sized hypothesis testing is essential in our upgraded CLR and retainment of our hypothesis test is defined as 0 $\in$ CR$_{\beta_w, 1-\alpha}$, we must ensure that this credible region has  $1-\alpha$ frequentist coverage. At low sample sizes the credible region is prone to being dominated by the prior. To mitigate this issue, we consider the PMP \citep{welch1963formulae} for the parameter of interest which was proven by \citet{datta1995priors} to match frequentist coverage to within $O(|{\mathcal{D}|}^{-1})$.

To employ the PMP in our setting, we do a preliminary step for computational simplicity: we first orthogonalize the treatment vector with respect to the observed covariates, i.e., $\Tilde{\w}_\mathcal{D} := (\I_{|\mathcal{D}|} - \Xdiscord(\Xdiscord^\top \Xdiscord)^{-1} \Xdiscord^\top)\w_\mathcal{D}$. This step ensures that the treatment vector is uncorrelated with the other covariates \citep{cox1987parameter}. Doing so also allows us to split the prior into (a) the uninformative PMP for the parameter of interest and (b) the informative prior for the nuisance parameters without any overlapping information \citep{tibshirani1989noninformative}. The noninformative PMP treatment effect has a Jeffreys-style prior proportional to the square root of its entry on the main diagonal of the Fisher information matrix for the discordant pairs. In the CLR setting, this value is computed as
\beqn
I_{ww}(\beta_w, \bbeta, \Xdiscord) = \sum_{i=1}^{|\mathcal{D}|}(\Tilde{w}_{{\mathcal{D}}_i})^2 p_i(1-p_i)
\eeqn
\noindent where $p_i$ is the likelihood of each pair respectively in \eqref{eq:discordant_pairs_clr} and thus a function of both $\beta_w, \bbeta$ and the two relevant rows of $\Xdiscord$. For the informative part (the prior or the nuisance parameters $\bbeta$), we will consider the naive MVN prior of~\eqref{eq:Normal_prior_without_intercept_flat_treatment}. Putting these two pieces together yields a prior proportional to $ 
\sqrt{I_{ww}(\beta_w, \bbeta, \Xdiscord)} \times [\text{Equation}~ \ref{eq:Normal_prior_without_intercept_flat_treatment}]$. 
%
%

\subsection{The Hybrid Prior: A Mixture of g's + PMP}\label{subsec:prior:gpmp}

The final prior we consider combines the advantages of the $g$-prior of and the PMP and thus is a prior proportional to  $\sqrt{I_{ww}(\beta_w, \bbeta, \Xdiscord)} \times [\text{Equation}~\ref{eq:g_prior}]$. 
%
%
\subsection{Our Software}\label{subsec:our_software}

Our BCLR flavors require a choice of model for the concordant pairs and a choice of the prior. Thus we have 12 flavors given by the combinations \{LR, GLMM, GEE\} $\times$ \{naive, $g$, PMP, $g+$PMP\}. Posterior inference for each flavor are implemented in our new \texttt{R} \citep{Rlang} package \ourpackage~on \texttt{CRAN}. Our package makes use of use \texttt{R}'s \texttt{Rcpp} package \citep{Rcpp2011} for its ability to optimize speed and \texttt{stan} \citep{stan2017} via \texttt{rstan} \citep{rstan2024} for its posterior sampling engine that implements Hamiltonian MCMC \citep{neal2011mcmc} with the no U-turn improvement \citep{hoffman2014no}. Our package \ourpackage~also has an option for the highest posterior density (HPD) inclusion test of \citet{box1973bayesian} for both contiguous and disjoint intervals using functionality from the \texttt{CODA} \citep{Plummer2006} and \texttt{ggdist} \citep{kay2024ggdist} packages respectively.  

\section{Numerical Studies}\label{sec:simulations}

\subsection{Synthetic Data}\label{subsec:synthetic}

We investigate performance of our BCLR methods and compare to four canonical methods appropriate in our setting: LR, CLR (both implemented in our package), GLMM (with a random intercept for each pair) implemented in \texttt{R}'s package \texttt{glmmTMB} \citep{glmmTMB2025} and GEE (with exchangeable correlation matrices for each pair) implemented in \texttt{R}'s package \texttt{geepack} \citep{geepack2025}. LR should do poorly as it ignores the dependence structure. For our method BCLR, we consider all 12 flavors (see Section~\ref{subsec:our_software}). We use the \texttt{rstan} defaults of random initialization, 1,000 burn-in iterations, 2,000 post-burn iterations and no thinning. We measure \qu{performance} as power when $\alpha = 5\%$. Power was estimated by the proportion of rejections of $H_0: \beta_w = 0$ where a rejection is defined as 0 $\notin$ CR$_{1 - \alpha}$ or 0 $\notin$ HPD$_{1 - \alpha}$.  

We examine $2n = 100, 250, 500$ and $p = 6$. Our $\X$ is generated via sampling a $n \times p$ matrix from $\uniform{-1}{1}$ and then duplicating each row with a small $\epsilon$ drawn from $\normnot{0}{0.05}$. We do this to simulate natural pairing environments, such as twin or pre-post studies, in which most covariates are nearly identical within each pair. The original rows and the duplicate ones are then zipped together such that row $\bv{x}_{i+1} = \bv{x}_{i} + \errorrv$. We then shuffle the first column so that it is uncorrelated within the pair, resulting in a non-trivial inference for $\beta_1$. This final $\X$ remains fixed throughout all simulation settings.

We then simulate the log-odds of positive response under a linear and nonlinear model. The linear model, $p_i = (1+\exp{-(\beta_0 + \bbeta \x_{i,\cdot} + \beta_w w_i)})^{-1}$, serves to assess performance without model misspecification nor covariate estimator bias. The parameter values we chose are $\beta_0 = -0.5$, $\bbeta = \bracks{1.25\,1.25\,\ldots}$. The non-linear logit-linked \citet{friedman1991multivariate} model,
\beqn
p_i = \inverse{1+\exp{-(\sin{\pi x_{i,1} x_{i,2}} + x_{i,3}^3 + x_{i,4}^2+x_{i,5}^2 + \beta_w w_i)}},
\eeqn
\noindent is designed to represent a more complex, realistic scenario. Here, fitting the linear model is merely a best linear approximation of the function above. In both probability response models, we chose $\beta_w = 0.5$, which was picked to reveal power differences among the methods for our range of considered sample sizes. (We additionally consider $\beta_w = 0$ to assess the size of the tests in each setting). 

We further consider two covariate settings where (1) only the first covariate is observed or (2) the first two covariates are observed. The setting of (1) simulates the case in which most of the predictive power lies in latent covariates that are virtually identical within pairs, while considering only a variable independent of the pair structure that has a significant effect on the response. The setting of (2) allows the model to peer into the underlying pair structure.

To reiterate, we have 3 (sample sizes)$\times$ 12+4 (our model flavors + canonical models) $\times$ 2 (power vs size) $\times$ 2 (covariate settings) $\times$ 2 (response models) = 384 settings. For each setting, we run $N_{sim} = 10,000$ simulation iterations. In each iteration, we draw $\w$ by uniformly setting each pair to either $(0,1)$ or $(1,0)$. Then we draw the $y_i$'s from independent Bernoullis with parameter $p_i$ given by the linear / nonlinear response models. Then we record the estimate of $\beta_w$ and whether the null hypothesis is rejected at a significance level of 5\%. Over the $N_{sim}$ simulations, we approximate power (when $\beta_w = 0.5$) and size (when $\beta_w = 0$) as the proportion of rejections.

For all 12 flavors of BCLR, we display comprehensive power and size results in Tables~\ref{tab:power_sub} and~\ref{tab:size_sub} respectively for the CR method. Tables~\ref{tab:power_sub_HPD} and \ref{tab:size_sub_HPD} in the supplementary information show power and size using the disjoint HPD testing method. The HPD testing method was found to have lower power. This is possibly due to our posteriors being skewed as HPD intervals overextend on the short tail. We also include posterior estimates of $g$ (for the $g$ and hybrid priors) in Tables~\ref{tab:g_values_0.5} and~\ref{tab:g_values_0} in the supplementary information for the cases of $\beta_w = 0.5$ and $\beta_w = 0$ respectively. Additionally, we include results on two more performance metrics in the supplementary information: the approximate mean squared error (MSE) in Table~\ref{tab:mse} and approximate coverage in Table~\ref{tab:coverage}. We use the posterior expectation (estimated by the average of the MCMC samples) as the definition of the point estimate $b_w$ and compute the squared error $(b_w - \beta_w)^2$. MSE is then approximated as the average of those squared errors over the $N_{sim}$ iterations. In each iteration we record if $\beta_w \in \text{CR}_{\beta_w, 95\%}$. The proportion of these set comparisons being true over the $N_{sim}$ iterations approximates the coverage.


\begin{table*}[!ht]
    \centering
    

\begin{subtable}{\textwidth}
    \centering
    \small
    \resizebox{\textwidth}{!}{%
    \begin{tabular}{ll llllll llllll}
        \multicolumn{2}{c}{} & \multicolumn{12}{c}{\textbf{True function}} \\
        \cmidrule(lr){3-14}
        \multicolumn{2}{c}{} & \multicolumn{6}{c}{linear} & \multicolumn{6}{c}{non-linear} \\ 
        \cmidrule(lr){3-8} \cmidrule(lr){9-14}
        \multicolumn{2}{c}{} & \multicolumn{12}{c}{Concordant pre-modeling step algorithm} \\
        \cmidrule(lr){3-14}
        & & \multicolumn{2}{c}{GEE} & \multicolumn{2}{c}{LR} & \multicolumn{2}{c}{GLMM} & \multicolumn{2}{c}{GEE} & \multicolumn{2}{c}{LR} & \multicolumn{2}{c}{GLMM} \\ 
        \cmidrule(lr){3-4} \cmidrule(lr){5-6} \cmidrule(lr){7-8} \cmidrule(lr){9-10} \cmidrule(lr){11-12} \cmidrule(lr){13-14}
        \textbf{N} & \textbf{Prior} & \textbf{one} & \textbf{two} & \textbf{one} & \textbf{two} & \textbf{one} & \textbf{two} & \textbf{one} & \textbf{two} & \textbf{one} & \textbf{two} & \textbf{one} & \textbf{two} \\ 
        \midrule
        
        \multirow{4}{*}{\textbf{100}}
          & Naive  & 0.1630  & 0.1627  & 0.1639  & 0.1631  & 0.1636  & 0.1620  & 0.2055* & 0.2045* & 0.2026* & 0.2114* & 0.2062* & 0.2128* \\
          & $g$    & 0.2011* & 0.1999* & 0.1996* & 0.1969* & 0.1970* & 0.1985* & 0.2122* & 0.2135* & 0.2121* & 0.2222* & 0.2196* & 0.2225* \\
          & PMP    & 0.1412  & 0.1401  & 0.1400  & 0.1367  & 0.1380  & 0.1385  & 0.1848* & 0.1845* & 0.1849  & 0.1886* & 0.1859* & 0.1867* \\
          & Hybrid & 0.1834* & 0.1864* & 0.1854* & 0.1827* & 0.1811* & 0.1814* & 0.1998* & 0.1967* & 0.1967* & 0.2043* & 0.2014* & 0.2004* \\ 
        \midrule
        \addlinespace
        
        \multirow{4}{*}{\textbf{250}}
          & Naive  & 0.3083  & 0.3065  & 0.3109  & 0.3185  & 0.3187  & 0.3176  & 0.3950  & 0.3942  & 0.3953  & 0.3844  & 0.3838  & 0.3822  \\
          & $g$    & 0.3371* & 0.3366* & 0.3366* & 0.3425* & 0.3409* & 0.3402* & 0.4008* & 0.4010* & 0.4022* & 0.3915* & 0.3887* & 0.3882* \\
          & PMP    & 0.2821* & 0.2829* & 0.2841  & 0.2888  & 0.2874  & 0.2875  & 0.3838  & 0.3843  & 0.3836  & 0.3698  & 0.3696  & 0.3682  \\
          & Hybrid & 0.3307* & 0.3298* & 0.3298* & 0.3297* & 0.3294* & 0.3319* & 0.3914* & 0.3924* & 0.3929* & 0.3788* & 0.3761  & 0.3782* \\
        \midrule
        \addlinespace
        
        \multirow{4}{*}{\textbf{500}}
          & Naive  & 0.5500  & 0.5517  & 0.5470  & 0.5424  & 0.5422  & 0.5412  & 0.6572  & 0.6571  & 0.6574  & 0.6531  & 0.6492  & 0.6496  \\
          & $g$    & 0.5684* & 0.5681* & 0.5705* & 0.5586  & 0.5562  & 0.5544  & 0.6619  & 0.6592  & 0.6575  & 0.6513  & 0.6501  & 0.6508  \\
          & PMP    & 0.5172  & 0.5168  & 0.5169  & 0.5057* & 0.5079* & 0.5060  & 0.6507  & 0.6501  & 0.6501  & 0.6431  & 0.6426  & 0.6417  \\
          & Hybrid & 0.5628* & 0.5630* & 0.5653* & 0.5504  & 0.5514  & 0.5483  & 0.6581  & 0.6580  & 0.6572  & 0.6478  & 0.6462  & 0.6454  \\
        \midrule
    \end{tabular}
    }
    \caption{\sf{Power analysis across different priors and pre-modeling algorithms.}}
    \label{tab:power_sub}
\end{subtable}


\begin{subtable}{\textwidth}
    \centering
    \small
    \resizebox{\textwidth}{!}{%
    \begin{tabular}{ll llllll llllll}
        \multicolumn{2}{c}{} & \multicolumn{12}{c}{\textbf{True function}} \\
        \cmidrule(lr){3-14}
        \multicolumn{2}{c}{} & \multicolumn{6}{c}{linear} & \multicolumn{6}{c}{non-linear} \\ 
        \cmidrule(lr){3-8} \cmidrule(lr){9-14}
        \multicolumn{2}{c}{} & \multicolumn{12}{c}{Concordant pre-modeling step algorithm} \\
        \cmidrule(lr){3-14}
        & & \multicolumn{2}{c}{GEE} & \multicolumn{2}{c}{LR} & \multicolumn{2}{c}{GLMM} & \multicolumn{2}{c}{GEE} & \multicolumn{2}{c}{LR} & \multicolumn{2}{c}{GLMM} \\ 
        \cmidrule(lr){3-4} \cmidrule(lr){5-6} \cmidrule(lr){7-8} \cmidrule(lr){9-10} \cmidrule(lr){11-12} \cmidrule(lr){13-14}
        \textbf{N} & \textbf{Prior} & \textbf{one} & \textbf{two} & \textbf{one} & \textbf{two} & \textbf{one} & \textbf{two} & \textbf{one} & \textbf{two} & \textbf{one} & \textbf{two} & \textbf{one} & \textbf{two} \\ 
        \midrule
        
        \multirow{4}{*}{\textbf{100}} 
          & Naive  & 0.0551  & 0.0566  & 0.0577  & 0.0574  & 0.0569  & 0.0574  & 0.0659* & 0.0645* & 0.0638* & 0.0748* & 0.0745* & 0.0743* \\
          & $g$    & 0.0813* & 0.0822* & 0.0825* & 0.0812* & 0.0804* & 0.0802* & 0.0749* & 0.0749* & 0.0762* & 0.0801* & 0.0821* & 0.0799* \\
          & PMP    & 0.0453  & 0.0465  & 0.0446  & 0.0510  & 0.0511  & 0.0502  & 0.0597* & 0.0613* & 0.0579  & 0.0683* & 0.0645* & 0.0677* \\
          & Hybrid & 0.0737* & 0.0732* & 0.0732* & 0.0738* & 0.0724* & 0.0726* & 0.0667* & 0.0669* & 0.0672* & 0.0746* & 0.0737* & 0.0729* \\
        \midrule
        \addlinespace
        
        \multirow{4}{*}{\textbf{250}} 
          & Naive  & 0.0474  & 0.0459  & 0.0461  & 0.0486  & 0.0484  & 0.0477  & 0.0576  & 0.0555  & 0.0574  & 0.0559  & 0.0559  & 0.0558  \\
          & $g$    & 0.0606* & 0.0616* & 0.0620* & 0.0653* & 0.0646* & 0.0628* & 0.0606* & 0.0625* & 0.0613* & 0.0600* & 0.0596* & 0.0618* \\
          & PMP    & 0.0411* & 0.0419* & 0.0425  & 0.0445  & 0.0456  & 0.0451  & 0.0571  & 0.0553  & 0.0559  & 0.0550  & 0.0538  & 0.0532  \\
          & Hybrid & 0.0587* & 0.0587* & 0.0582* & 0.0610* & 0.0604* & 0.0615* & 0.0608* & 0.0583* & 0.0603* & 0.0583* & 0.0561  & 0.0593* \\
        \midrule
        \addlinespace
        
        \multirow{4}{*}{\textbf{500}} 
          & Naive  & 0.0480  & 0.0480  & 0.0477  & 0.0447  & 0.0437  & 0.0446  & 0.0532  & 0.0536  & 0.0534  & 0.0534  & 0.0550  & 0.0540  \\
          & $g$    & 0.0640* & 0.0621* & 0.0632* & 0.0563  & 0.0555  & 0.0570  & 0.0548  & 0.0565  & 0.0547  & 0.0561  & 0.0578  & 0.0560  \\
          & PMP    & 0.0449  & 0.0447  & 0.0447  & 0.0418* & 0.0419* & 0.0426  & 0.0501  & 0.0532  & 0.0508  & 0.0545  & 0.0546  & 0.0544  \\
          & Hybrid & 0.0646* & 0.0603* & 0.0604* & 0.0539  & 0.0533  & 0.0551  & 0.0544  & 0.0534  & 0.0539  & 0.0569  & 0.0555  & 0.0577  \\ 
        \midrule
    \end{tabular}
    }
    \caption{\sf{Size analysis across different priors and pre-modeling algorithms.}}
    \label{tab:size_sub}
\end{subtable}
    \caption{Side-by-side analysis of simulated power and size (to four decimals) approximated over the $N_{sim} = 10,000$ iterations for all flavors of BCLR. Asterisks in the size table indicate a rejection of the null that the size equals 5\% using a binomial exact test at the Bonferroni-corrected family-wise error rate of 5\% within the context of this display. Asterisks in the power table mirror the size table to indicate the test size is suspect.}
    \label{tab:power_and_size_sub}
\end{table*}

To simplify the results illustration in the manuscript, we will display only the combination choice we believe is \qu{best} as determined by (1) being mostly properly sized across the various simulation settings and (2) being numerically stable. The winning flavor we believe to be LR+naive and henceforth termed BCLR in this section's results henceforth. Although GEE+$g$ had the highest power on average over our settings, both GEE and $g$ we believe have separate problems which prevents us from recommending it to a practitioner. GEE suffered from occasional unacceptably large MSE. In some of our iterations, a covariate perfectly predicts the outcome. Here, the iterative reweighted least squares algorithm will fail to converge as the parameter estimates run towards infinity. Additionally, the sandwich estimator used for robust standard errors in GEE can produce singular matrices. In our software implementation, we have many types of robust fallbacks to LR. GLMM also experienced such instabilities, but less so. The $g$ prior (and hence the hybrid as well) experiences inflated size. This is likely due to $g$ being large, yielding flat priors. The PMP is observed to be improperly sized being both too small and too large. Although the naive prior cannot weight the concordant data explicitly like in a $g$ prior (a weighting we conjectured at the outset to be critical), the sample size $|\mathcal{C}|$ is built implicitly into $\bSigmaconcord$, its variance estimate.

The power and size results for BCLR and the canonical models are found in Table~\ref{tab:power_size_vs_competitors} for each sample size $\times$ covariate setting $\times$ response model. BCLR's power performance dominated CLR's power performance in all settings. BCLR is also on par with the canonical methods GEE and GLMM in the linear logodds response setting. However, in the nonlinear log odds response setting, BCLR dominates GEE and GLMM, methods that rely heavily on the linear model assumption. BCLR uses matched pair covariate data differences as seen in \eqref{eq:discordant_pairs_clr} which have a long history of being robust in nonlinear response models. These differences reduce bias because the nonlinearities \qu{cancel out} or are minimized \citep{rubin1973, ho2007matching}. As expected LR underperforms as it is misspecified.

\begin{table*}[ht]
    \centering
    \setlength{\tabcolsep}{4pt} 

    \begin{tabular}{ll c cccc c cccc}
        & & & \multicolumn{4}{c}{\textbf{True function (Power)}} & & \multicolumn{4}{c}{\textbf{True function (Size)}} \\ 
        \cmidrule(lr){4-7} \cmidrule(lr){9-12}
        & & & \multicolumn{2}{c}{linear} & \multicolumn{2}{c}{non-linear} & & \multicolumn{2}{c}{linear} & \multicolumn{2}{c}{non-linear} \\ 
        \cmidrule(lr){4-5} \cmidrule(lr){6-7} \cmidrule(lr){9-10} \cmidrule(lr){11-12}
        \textbf{N} & \textbf{Model} & & \multicolumn{2}{c}{Covariates} & \multicolumn{2}{c}{Covariates} & & \multicolumn{2}{c}{Covariates} & \multicolumn{2}{c}{Covariates} \\
        & & & \textbf{one} & \textbf{two} & \textbf{one} & \textbf{two} & & \textbf{one} & \textbf{two} & \textbf{one} & \textbf{two} \\ 
        \midrule
        
        \multirow{5}{*}{\textbf{100}}
        & BCLR & & 0.1683 & 0.1701 & 0.2143* & 0.2152* & & 0.0552 & 0.0556 & 0.0679* & 0.0674* \\ 
        & CLR  & & 0.0780 & 0.0599 & 0.1492 & 0.1359 & & 0.0161* & 0.0143* & 0.0418* & 0.0418* \\ 
        & GEE  & & 0.1757 & 0.1756 & 0.1880 & 0.1882 & & 0.0553 & 0.0546 & 0.0537 & 0.0537 \\ 
        & GLMM & & 0.1145 & 0.1191 & 0.1704 & 0.1722 & & 0.0282* & 0.0295* & 0.0459 & 0.0464 \\ 
        & LR   & & 0.0886 & 0.1026 & 0.1670 & 0.1695 & & 0.0195* & 0.0232* & 0.0442 & 0.0447 \\ 
        \addlinespace 
        
        \toprule
        
        \multirow{5}{*}{\textbf{250}}
        & BCLR & & 0.3172 & 0.3202 & 0.3909 & 0.3894 & & 0.0469 & 0.0477 & 0.0573 & 0.0569 \\ 
        & CLR  & & 0.2961 & 0.2974 & 0.3616 & 0.3614 & & 0.0431 & 0.0447 & 0.0478 & 0.0494 \\ 
        & GEE  & & 0.3513 & 0.3533 & 0.3737 & 0.3739 & & 0.0497 & 0.0491 & 0.0503 & 0.0505 \\ 
        & GLMM & & 0.3028 & 0.3071 & 0.3652 & 0.3655 & & 0.0369* & 0.0375* & 0.0464 & 0.0465 \\ 
        & LR   & & 0.2189 & 0.2417 & 0.3562 & 0.3562 & & 0.0184* & 0.0219* & 0.0440 & 0.0444 \\ 
        \addlinespace
        
        \toprule
        \multirow{5}{*}{\textbf{500}}
        & BCLR & & 0.5531 & 0.5523 & 0.6517 & 0.6526 & & 0.0465 & 0.0467 & 0.0567 & 0.0570 \\ 
        & CLR  & & 0.5420 & 0.5415 & 0.6361 & 0.6344 & & 0.0490 & 0.0505 & 0.0504 & 0.0515 \\ 
        & GEE  & & 0.6045 & 0.6039 & 0.6422 & 0.6420 & & 0.0540 & 0.0530 & 0.0511 & 0.0510 \\ 
        & GLMM & & 0.5656 & 0.5676 & 0.6347 & 0.6349 & & 0.0410* & 0.0418* & 0.0472 & 0.0474 \\ 
        & LR   & & 0.4480 & 0.4793 & 0.6235 & 0.6241 & & 0.0187* & 0.0233* & 0.0442 & 0.0443 \\ 
        \midrule
    \end{tabular}

    \caption{Analysis of simulated power and size for BCLR and canonical methods ($N_{sim} = 10,000$). Asterisks follow the same convention as in Table~\ref{tab:power_and_size_sub}.}
    \label{tab:power_size_vs_competitors}
\end{table*}

\subsection{Real Data Application}\label{subsec:real_data}

The Framingham Heart Study \citep[FHS,][]{dawber1951epidemiological, tsao2015cohort} is one of the most significant and longest-running longitudinal studies in medical history, dating to 1948 with its initial $\approx$5,000 adult subjects. We make use of the study's publicly available dataset as found in the \texttt{R} package \texttt{riskCommunicator} \citep{Grembi20222}. 

There are many endpoints in this study. We use the binary endpoint \qu{prevalent coronary heart disease} where the positive level was defined as preexisting angina pectoris, myocardial infarction (hospitalized, silent or unrecognized), or coronary insufficiency (unstable angina). We considered eight covariates: total cholesterol (numeric), systolic blood pressure (numeric), diastolic blood pressure (numeric), heart rate (numeric), number of cigarettes smoked per day (count), body mass index (numeric), diabetes (binary), and use of anti-hypertensive medication (binary). These covariates varied longitudinally across patients and had low missingness. We defined a pair as the first measurement ($w = 0$) and the last measurement ($w = 1$) made on the same patient. Intermediate measurements were dropped. We then dropped missing data and retained only the patients who had these two measurements. For these patients, the time between first and last measurement was always 10-14 years elapsed, and 82\% of the subjects had 12 years elapsed. Hence, the treatment effect of interest is the additive effect of approximately 12 years of aging on the log-odds of a prevalent coronary heart disease event. As aging increases the risk of heart disease, the treatment effect is real and positive.

After this initial data curation, there were $n=2,971$ patient pairs. In the approximate 12yr elapse, most patients' endpoints remained the same; there were $|\mathcal{C}|$ = 2,718 concordant pairs and only $|\mathcal{D}|$ = 253 discordant pairs. Hence, this is an example where CLR's performance is likely to suffer as 91\% of the subject data is discarded. Conversely, this is an example where our BCLR will perform well as we use that 91\% of the data to estimate accurately the approximate linear effect of the eight informative covariates on the log odds of prevalent coronary heart disease. This is exactly what is found in the observed estimates and significant tests (see Table~\ref{tab:real_world_results}) where CLR failed to find the real effect and BCLR comfortably found the effect. As an analogy, in the FHS data BCLR is using covariate adjustment to find a noisy effect while CLR is using a model without covariates at all.

\begin{table}[ht]
\centering
\begin{tabular}{lccc}
     & estimate & $\beta_w$ estimator & Approximate $p_{val}$  \\ 
    Method & of $\beta_w$ &  variance & for $H_0: \beta_w = 0$ \\     
    \midrule
    CLR  & 26.57 & 39800 & 0.9995 \\
    BCLR & 7.54  & 2.06  & 0.0000 \\
        \midrule 
\end{tabular}
\caption{Comparison of CLR and BCLR inference for the FHS data.}
\label{tab:real_world_results}
\end{table}

\section{Discussion}\label{sec:discussion}

We developed an improvement to the classic CLR model in the setting of estimating a binary treatment effect along with control covariates. The source of our improvement is the information-salvaging of the concordant pairs which are traditionally viewed as ancillary to the likelihood. Our method BCLR is a two-step procedure: (1) we use the information about the control covariates found in the concordant pairs by creating a pre-model, and (2) we use this information as a prior in a Bayesian implementation of CLR, which we call BCLR. Thus we provide a principled way to recover efficiency that is lost in the standard workflow. 

We investigated many different flavors of pre-models, but we found through simulation that the most reliable was the prosaic LR whose coefficient estimates and variance-covariance matrix of the coefficients' estimator were rolled into a prosaic multivariate normal prior. By \qu{most reliable} we mean it provided tests of the parameter of interest that were properly sized and protected from numeric instability, which is a risk due to the structure of the concordant pair data. 

Our BCLR method was found to be superior to CLR as measured by the power of the test of no treatment effect, and it was more numerically stable. (Our method also translates to improvements for inference on the nuisance parameters, but this is unshown herein). The BCLR method was also competitive with other canonical models, such as GEE and GLMM, although GEE seems to perform better. In the case of a non-linear log odds of positive response model, BCLR was found superior. We hesitate to state numerically how much better BCLR's power is over the classic CLR as improvements are setting-dependent. But we can say that in the simulated synthetic data settings of Section~\ref{subsec:synthetic} where we examined both linear and nonlinear log-odds of positive response models, improvements ranged between 58-116\% for $n=100$, between 7.1-8.1\% for $n=250$ and between 2.0-2.9\% for $n=500$. Additionally, in the case of a real dataset (the FHS of Section~\ref{subsec:real_data}), BCLR was able to decisively identify a known effect while CLR could not.

As for GEE performing better than BCLR in the linear case, we do not believe this to be an apples-to-apples comparison as GEE provides estimates of the population-level effects, thus answering the question, \qu{What is the change in the log odds of positive response for the whole population}? where BCLR provides estimates of subject-specific effects, thus answering the question, \qu{What is the change in the log odds of positive response for the average specific subject}? and hence the parameters are defined differently.

We have many further paths to explore. We believe our methods could be extended to strata with more than two subjects, continuously-measured treatments, and categorical treatments, and it is our plan to update the software to make such accommodations. We would also like to explore the flavors more, e.g., we can correct the large MSE due to numerical instability in GEE and GLMM by reverting to LR for those few cases. There are also other flavors worth exploring such as power priors  \citep{ibrahim2000power}. Additionally, we believe we can prove the finite efficiency of BCLR vs CLR for the estimation of the additive treatment effect parameter. This would involve making assumptions on the distribution of covariates and the distribution of the parameters $\bbeta$. Then we would need to carefully compute the variance of the naive+LR estimator (and possibly other flavors), which is likely doable.
\vspace{0cm}

\section{Conflicts of interest}
The authors declare that they have no competing interests.

\section{Funding}
This research was supported by Grant No 2018112 from the United States-Israel Binational Science Foundation (BSF).

\section{Data availability}
The data and code to reproduce the results herein can be found at \url{https://github.com/Tennenbaum-J/bclogit_package_and_paper_repo/}.

\section{Author contributions statement}

J.T. and A.K. shared in conception and manuscript writing and review. J.T. did the majority of the numerical studies. J.T and A.K. shared in the development of the \texttt{CRAN} package \ourpackage.

\section{Acknowledgments}
The authors thank the support of Grant No 2018112 from the United States-Israel Binational Science Foundation (BSF). We thank the Gemini CLI for aiding with software development. 


\bibliographystyle{oup-abbrvnat}
\bibliography{refs}

\clearpage

\begin{appendices}
\onecolumn
\renewcommand{\thefigure}{A\arabic{figure}}
\setcounter{figure}{0}
\renewcommand{\thetable}{A\arabic{table}}
\setcounter{table}{0}

\begin{center}
    {\large Supplementary information for \\
    \qu{\ourtitle}} \\~\\
    \normalsize by Jacob Tennenbaum and Adam Kapelner\\
    
\end{center}
\vspace{-0.5cm}

\begin{table*}[!ht]
    \centering
    

\begin{subtable}{\textwidth}
    \centering
    \small
    \resizebox{\textwidth}{!}{%
    \begin{tabular}{ll llllll llllll}
        \multicolumn{2}{c}{} & \multicolumn{12}{c}{\textbf{True function}} \\
        \cmidrule(lr){3-14}
        \multicolumn{2}{c}{} & \multicolumn{6}{c}{linear} & \multicolumn{6}{c}{non-linear} \\ 
        \cmidrule(lr){3-8} \cmidrule(lr){9-14}
        \multicolumn{2}{c}{} & \multicolumn{12}{c}{Concordant pre-modeling step algorithm} \\
        \cmidrule(lr){3-14}
        & & \multicolumn{2}{c}{GEE} & \multicolumn{2}{c}{LR} & \multicolumn{2}{c}{GLMM} & \multicolumn{2}{c}{GEE} & \multicolumn{2}{c}{LR} & \multicolumn{2}{c}{GLMM} \\ 
        \cmidrule(lr){3-4} \cmidrule(lr){5-6} \cmidrule(lr){7-8} \cmidrule(lr){9-10} \cmidrule(lr){11-12} \cmidrule(lr){13-14}
        \textbf{N} & \textbf{Prior} & \textbf{one} & \textbf{two} & \textbf{one} & \textbf{two} & \textbf{one} & \textbf{two} & \textbf{one} & \textbf{two} & \textbf{one} & \textbf{two} & \textbf{one} & \textbf{two} \\ 
        \midrule
        
        \multirow{4}{*}{\textbf{100}}
          & Naive  & 0.1492  & 0.1492  & 0.1474  & 0.1488  & 0.1491  & 0.1485  & 0.1888* & 0.1891* & 0.1874  & 0.1941* & 0.1929* & 0.1969* \\
          & $g$    & 0.1658* & 0.1657* & 0.1665* & 0.1624* & 0.1635* & 0.1625* & 0.1965* & 0.1946* & 0.1940* & 0.2017* & 0.1992* & 0.2020* \\
          & PMP    & 0.1276* & 0.1249* & 0.1261* & 0.1195  & 0.1222* & 0.1221* & 0.1701  & 0.1682  & 0.1697  & 0.1697* & 0.1684  & 0.1666* \\
          & Hybrid & 0.1486  & 0.1481  & 0.1477  & 0.1407  & 0.1400  & 0.1431  & 0.1832* & 0.1813  & 0.1784  & 0.1837* & 0.1831* & 0.1813* \\
        \midrule
        \addlinespace
        
        \multirow{4}{*}{\textbf{250}}
          & Naive  & 0.3040  & 0.3061  & 0.3065  & 0.3161  & 0.3136  & 0.3113  & 0.3912  & 0.3898  & 0.3928  & 0.3800  & 0.3769  & 0.3782  \\
          & $g$    & 0.3271* & 0.3246  & 0.3238* & 0.3295* & 0.3307* & 0.3288* & 0.3947* & 0.3963* & 0.3953* & 0.3836* & 0.3817* & 0.3829* \\
          & PMP    & 0.2800* & 0.2753* & 0.2814* & 0.2825  & 0.2847  & 0.2817  & 0.3784  & 0.3769  & 0.3800  & 0.3642  & 0.3673  & 0.3621  \\
          & Hybrid & 0.3218  & 0.3148  & 0.3195  & 0.3193  & 0.3197  & 0.3206  & 0.3854* & 0.3897* & 0.3868* & 0.3717  & 0.3712  & 0.3727  \\
        \midrule
        \addlinespace
        
        \multirow{4}{*}{\textbf{500}}
          & Naive  & 0.5499  & 0.5483  & 0.5484  & 0.5405  & 0.5417  & 0.5408  & 0.6586  & 0.6571  & 0.6552  & 0.6497  & 0.6502  & 0.6501  \\
          & $g$    & 0.5666* & 0.5653* & 0.5669* & 0.5566  & 0.5540  & 0.5530  & 0.6642  & 0.6586  & 0.6599  & 0.6521  & 0.6544* & 0.6529  \\
          & PMP    & 0.5195  & 0.5161  & 0.5130  & 0.5053* & 0.5063* & 0.5051* & 0.6495  & 0.6510  & 0.6495  & 0.6432  & 0.6405  & 0.6426  \\
          & Hybrid & 0.5611* & 0.5599* & 0.5626* & 0.5501  & 0.5473  & 0.5447  & 0.6587  & 0.6611  & 0.6556  & 0.6466  & 0.6474  & 0.6457* \\
        \midrule
    \end{tabular}
    }
    \caption{\sf{Power analysis across different priors and pre-modeling algorithms.}}
    \label{tab:power_sub_HPD}
\end{subtable}


\begin{subtable}{\textwidth}
    \centering
    \small
    \resizebox{\textwidth}{!}{%
    \begin{tabular}{ll llllll llllll}
        \multicolumn{2}{c}{} & \multicolumn{12}{c}{\textbf{True function}} \\
        \cmidrule(lr){3-14}
        \multicolumn{2}{c}{} & \multicolumn{6}{c}{linear} & \multicolumn{6}{c}{non-linear} \\ 
        \cmidrule(lr){3-8} \cmidrule(lr){9-14}
        \multicolumn{2}{c}{} & \multicolumn{12}{c}{Concordant pre-modeling step algorithm} \\
        \cmidrule(lr){3-14}
        & & \multicolumn{2}{c}{GEE} & \multicolumn{2}{c}{LR} & \multicolumn{2}{c}{GLMM} & \multicolumn{2}{c}{GEE} & \multicolumn{2}{c}{LR} & \multicolumn{2}{c}{GLMM} \\ 
        \cmidrule(lr){3-4} \cmidrule(lr){5-6} \cmidrule(lr){7-8} \cmidrule(lr){9-10} \cmidrule(lr){11-12} \cmidrule(lr){13-14}
        \textbf{N} & \textbf{Prior} & \textbf{one} & \textbf{two} & \textbf{one} & \textbf{two} & \textbf{one} & \textbf{two} & \textbf{one} & \textbf{two} & \textbf{one} & \textbf{two} & \textbf{one} & \textbf{two} \\ 
        \midrule
        
        \multirow{4}{*}{\textbf{100}} 
          & Naive  & 0.0486  & 0.0491  & 0.0512  & 0.0502  & 0.0494  & 0.0513  & 0.0589* & 0.0597* & 0.0579  & 0.0697* & 0.0664* & 0.0690* \\
          & $g$    & 0.0618* & 0.0600* & 0.0596* & 0.0601* & 0.0588* & 0.0583* & 0.0640* & 0.0651* & 0.0653* & 0.0727* & 0.0720* & 0.0708* \\
          & PMP    & 0.0386* & 0.0395* & 0.0379* & 0.0424  & 0.0422* & 0.0420* & 0.0539  & 0.0528  & 0.0514  & 0.0595* & 0.0577  & 0.0600* \\
          & Hybrid & 0.0487  & 0.0496  & 0.0513  & 0.0512  & 0.0500  & 0.0488  & 0.0589* & 0.0577  & 0.0577  & 0.0646* & 0.0629* & 0.0617* \\
        \midrule
        \addlinespace
        
        \multirow{4}{*}{\textbf{250}} 
          & Naive  & 0.0457  & 0.0467  & 0.0444  & 0.0476  & 0.0471  & 0.0481  & 0.0567  & 0.0565  & 0.0566  & 0.0574  & 0.0563  & 0.0566  \\
          & $g$    & 0.0582* & 0.0571  & 0.0592* & 0.0600* & 0.0594* & 0.0589* & 0.0601* & 0.0617* & 0.0604* & 0.0611* & 0.0593* & 0.0623* \\
          & PMP    & 0.0407* & 0.0404* & 0.0413* & 0.0439  & 0.0433  & 0.0449  & 0.0547  & 0.0543  & 0.0568  & 0.0540  & 0.0544  & 0.0540  \\
          & Hybrid & 0.0538  & 0.0554  & 0.0561  & 0.0571  & 0.0566  & 0.0555  & 0.0590* & 0.0600* & 0.0594* & 0.0565  & 0.0548  & 0.0569  \\
        \midrule
        \addlinespace
        
        \multirow{4}{*}{\textbf{500}} 
          & Naive  & 0.0492  & 0.0479  & 0.0461  & 0.0460  & 0.0458  & 0.0472  & 0.0534  & 0.0527  & 0.0555  & 0.0561  & 0.0555  & 0.0552  \\
          & $g$    & 0.0619* & 0.0622* & 0.0636* & 0.0553  & 0.0550  & 0.0548  & 0.0553  & 0.0576  & 0.0570  & 0.0570  & 0.0583* & 0.0555  \\
          & PMP    & 0.0449  & 0.0453  & 0.0456  & 0.0420* & 0.0421* & 0.0422* & 0.0520  & 0.0530  & 0.0525  & 0.0557  & 0.0548  & 0.0567  \\
          & Hybrid & 0.0605* & 0.0600* & 0.0617* & 0.0533  & 0.0533  & 0.0545  & 0.0551  & 0.0553  & 0.0567  & 0.0566  & 0.0578  & 0.0597* \\
        \midrule
    \end{tabular}
    }
    \caption{\sf{Size analysis across different priors and pre-modeling algorithms.}}
    \label{tab:size_sub_HPD}
\end{subtable}
    \caption{Side-by-side analysis of simulated power and size (to four decimals) approximated over the $N_{sim} = 10,000$ iterations for all flavors of BCLR for the disjoint HPD testing method. Asterisks in the size table indicate a rejection of the null that the size equals 5\% using a binomial exact test at the Bonferroni-corrected family-wise error rate of 5\% within the context of this display. Asterisks in the power table mirror the size table to indicate the test size is suspect.}
\end{table*}

\begin{table*}[ht]
    \centering

    \begin{subtable}{\textwidth}
        \centering
        \resizebox{\textwidth}{!}{%
        \begin{tabular}{ll cccccc cccccc}
            \multicolumn{2}{c}{} & \multicolumn{12}{c}{\textbf{True function}} \\
            \cmidrule(lr){3-14}
            \multicolumn{2}{c}{} & \multicolumn{6}{c}{linear} & \multicolumn{6}{c}{non-linear} \\ 
            \cmidrule(lr){3-8} \cmidrule(lr){9-14}
            \multicolumn{2}{c}{} & \multicolumn{12}{c}{Concordant pre-modeling step algorithm} \\
            \cmidrule(lr){3-14}
            & & \multicolumn{2}{c}{GEE} & \multicolumn{2}{c}{LR} & \multicolumn{2}{c}{GLMM} & \multicolumn{2}{c}{GEE} & \multicolumn{2}{c}{LR} & \multicolumn{2}{c}{GLMM} \\ 
            \cmidrule(lr){3-4} \cmidrule(lr){5-6} \cmidrule(lr){7-8} \cmidrule(lr){9-10} \cmidrule(lr){11-12} \cmidrule(lr){13-14}
            \textbf{N} & \textbf{Prior} & \textbf{one} & \textbf{two} & \textbf{one} & \textbf{two} & \textbf{one} & \textbf{two} & \textbf{one} & \textbf{two} & \textbf{one} & \textbf{two} & \textbf{one} & \textbf{two} \\ 
            \midrule

            \multirow{2}{*}{\textbf{100}}
              & $g$    & 5.3578 & 5.8599 & 7.0569 & 5.2688 & 8.1952 & 7.0824 & 2.2851 & 4.0994 & 1.7775 & 2.7343 & 1.9395 & 2.7363 \\
              & Hybrid & 2.6820 & 2.5799 & 2.5692 & 2.8141 & 2.5423 & 2.5416 & 2.3493 & 3.1161 & 1.7694 & 2.0428 & 1.9772 & 2.1401 \\
            \midrule
            \addlinespace

            \multirow{2}{*}{\textbf{250}}
              & $g$    & 2.1917 & 2.1326 & 2.1829 & 2.1228 & 2.1678 & 2.1193 & 2.1286 & 2.0498 & 2.1445 & 2.0692 & 2.1383 & 2.0987 \\
              & Hybrid & 2.1692 & 2.1183 & 2.1623 & 2.1504 & 2.1731 & 2.1252 & 2.1164 & 2.0560 & 2.1310 & 2.0493 & 2.1331 & 2.0579 \\
            \midrule
            \addlinespace

            \multirow{2}{*}{\textbf{500}} 
              & $g$    & 2.4078 & 2.3235 & 2.4239 & 2.3282 & 2.4146 & 2.3346 & 2.4096 & 2.2724 & 2.4177 & 2.2777 & 2.4274 & 2.2718 \\
              & Hybrid & 2.4093 & 2.3348 & 2.4010 & 2.3300 & 2.4088 & 2.3360 & 2.4280 & 2.2786 & 2.4336 & 2.2849 & 2.4175 & 2.2740 \\
            \midrule
        \end{tabular}
        }
        \caption{$\beta_w = 0.5$}
        \label{tab:g_values_0.5}
    \end{subtable}
    
    \vspace{1.5em} 
    
    \begin{subtable}{\textwidth}
        \centering
        \resizebox{\textwidth}{!}{%
        \begin{tabular}{ll cccccc cccccc}
            \multicolumn{2}{c}{} & \multicolumn{12}{c}{\textbf{True function}} \\
            \cmidrule(lr){3-14}
            \multicolumn{2}{c}{} & \multicolumn{6}{c}{linear} & \multicolumn{6}{c}{non-linear} \\ 
            \cmidrule(lr){3-8} \cmidrule(lr){9-14}
            \multicolumn{2}{c}{} & \multicolumn{12}{c}{Concordant pre-modeling step algorithm} \\
            \cmidrule(lr){3-14}
            & & \multicolumn{2}{c}{GEE} & \multicolumn{2}{c}{LR} & \multicolumn{2}{c}{GLMM} & \multicolumn{2}{c}{GEE} & \multicolumn{2}{c}{LR} & \multicolumn{2}{c}{GLMM} \\ 
            \cmidrule(lr){3-4} \cmidrule(lr){5-6} \cmidrule(lr){7-8} \cmidrule(lr){9-10} \cmidrule(lr){11-12} \cmidrule(lr){13-14}
            \textbf{N} & \textbf{Prior} & \textbf{one} & \textbf{two} & \textbf{one} & \textbf{two} & \textbf{one} & \textbf{two} & \textbf{one} & \textbf{two} & \textbf{one} & \textbf{two} & \textbf{one} & \textbf{two} \\ 
            \midrule

            \multirow{2}{*}{\textbf{100}}
              & $g$    & 5.7017 & 5.7392 & 5.3346 & 4.4320 & 5.2387 & 4.8964 & 1.7905 & 2.7523 & 1.8032 & 1.8968 & 1.8017 & 2.0032 \\
              & Hybrid & 2.7943 & 2.5556 & 2.5261 & 2.6211 & 2.6957 & 2.8051 & 1.7848 & 2.8216 & 1.7959 & 1.8437 & 1.7842 & 2.0088 \\
              \midrule
            \addlinespace

            \multirow{2}{*}{\textbf{250}}
              & $g$    & 2.1778 & 2.1364 & 2.1299 & 2.1477 & 2.1520 & 2.1610 & 2.1403 & 2.0715 & 2.1463 & 2.0763 & 2.1463 & 2.0838 \\
              & Hybrid & 2.1876 & 2.1187 & 2.1684 & 2.1285 & 2.1197 & 2.1009 & 2.1575 & 2.0805 & 2.1868 & 2.0801 & 2.1511 & 2.0694 \\
              \midrule
            \addlinespace

            \multirow{2}{*}{\textbf{500}} 
              & $g$    & 2.3939 & 2.3036 & 2.4084 & 2.3080 & 2.3949 & 2.3070 & 2.4408 & 2.2976 & 2.4349 & 2.2923 & 2.4567 & 2.2971 \\
              & Hybrid & 2.4340 & 2.3033 & 2.4124 & 2.3324 & 2.4006 & 2.3132 & 2.4532 & 2.2991 & 2.4652 & 2.2906 & 2.4460 & 2.3065 \\
            \midrule
        \end{tabular}
        }
        \caption{$\beta_w = 0$}
        \label{tab:g_values_0}
    \end{subtable}
    \caption{$\log_{10}(g)$ (to four decimal places) for relevant priors averaged over the $N_{sim} = 10,000$ iterations.}
    \label{tab:g_values_combined}

\end{table*}

\begin{table*}[!ht]
\centering
\scriptsize
\resizebox{6in}{!}{%
\begin{tabular}{ll cccc cccc}
\multicolumn{2}{c}{} & \multicolumn{8}{c}{\textbf{Treatment effect}} \\ 
\cmidrule(lr){3-10}
\multicolumn{2}{c}{} & \multicolumn{4}{c}{$\beta_w = 0$} & \multicolumn{4}{c}{$\beta_w = 0.5$} \\ 
\cmidrule(lr){3-6} \cmidrule(lr){7-10}
\multicolumn{2}{c}{} & \multicolumn{2}{c}{linear} & \multicolumn{2}{c}{non-linear} & \multicolumn{2}{c}{linear} & \multicolumn{2}{c}{non-linear} \\ 
\cmidrule(lr){3-4} \cmidrule(lr){5-6} \cmidrule(lr){7-8} \cmidrule(lr){9-10}
\textbf{N} & \textbf{Model} & \textbf{one} & \textbf{two} & \textbf{one} & \textbf{two} & \textbf{one} & \textbf{two} & \textbf{one} & \textbf{two} \\ 
\midrule

\multirow{16}{*}{\textbf{100}}
 & BCLR Naive $\times$ GEE      & 0.3471 & 0.3654 & 0.2605 & 0.2894 & 0.3621 & 0.3788 & 0.3205 & 6.5415 \\
 & BCLR Naive $\times$ LR       & 0.3467 & 0.3635 & 0.2597 & 0.2738 & 0.3620 & 0.3782 & 0.3169 & 0.3289 \\
 & BCLR Naive $\times$ GLMM     & 0.3466 & 0.3653 & 0.2598 & 0.2782 & 0.3614 & 0.3783 & 0.3180 & 2.0188 \\
 & BCLR $g$ prior $\times$ GEE  & 0.9495 & 0.9692 & 0.3172 & 0.3297 & 1.1249 & 1.1675 & 0.4304 & 0.4539 \\
 & BCLR $g$ prior $\times$ LR   & 0.9435 & 0.9704 & 0.3176 & 0.3305 & 1.1211 & 1.1659 & 0.4318 & 0.4555 \\
 & BCLR $g$ prior $\times$ GLMM & 0.9571 & 0.9725 & 0.3175 & 0.3290 & 1.1291 & 1.1677 & 0.4305 & 0.4561 \\
 & BCLR PMP $\times$ GEE        & 0.3232 & 0.3790 & 0.2492 & 0.2972 & 0.3284 & 0.3916 & 0.3039 & 4.3445 \\
 & BCLR PMP $\times$ LR         & 0.3230 & 0.3777 & 0.2488 & 0.2780 & 0.3271 & 0.3903 & 0.2999 & 0.3363 \\
 & BCLR PMP $\times$ GLMM       & 0.3230 & 0.3803 & 0.2487 & 0.2837 & 0.3272 & 0.3911 & 0.3023 & 0.8877 \\
 & BCLR Hybrid $\times$ GEE     & 1.1783 & 1.2654 & 0.2879 & 0.3219 & 1.5050 & 1.7191 & 0.3893 & 0.4524 \\
 & BCLR Hybrid $\times$ LR      & 1.0497 & 1.2842 & 0.2875 & 0.3221 & 1.3965 & 1.6835 & 0.3868 & 0.4420 \\
 & BCLR Hybrid $\times$ GLMM    & 1.0493 & 1.1984 & 0.2879 & 0.3225 & 1.4814 & 1.5379 & 0.3875 & 0.4390 \\
 & CLR                          & 256.75 & 6066.5 & 0.2601 & 0.3515 & 418.08 & 4810.2 & 3.0314 & 88.891 \\
 & GEE                          & 0.1288 & 0.1511 & 0.1756 & 0.1797 & 0.1500 & 0.1651 & 0.2035 & 0.2080 \\
 & GLMM                         & 0.4872 & 0.7979 & 0.2084 & 0.2110 & 0.8329 & 1.6927 & 0.2499 & 0.2528 \\
 & LR                           & 0.1291 & 0.1515 & 0.1755 & 0.1796 & 0.1502 & 0.1654 & 0.2035 & 0.2079 \\ 
\midrule

\multirow{16}{*}{\textbf{250}}
 & BCLR Naive $\times$ GEE      & 0.1076 & 0.1115 & 0.0838 & 0.0832 & 0.1100 & 0.1154 & 0.0954 & 0.0951 \\
 & BCLR Naive $\times$ LR       & 0.1075 & 0.1116 & 0.0839 & 0.0828 & 0.1099 & 0.1153 & 0.0954 & 0.0951 \\
 & BCLR Naive $\times$ GLMM     & 0.1077 & 0.1117 & 0.0838 & 0.0827 & 0.1099 & 0.1152 & 0.0954 & 0.0949 \\
 & BCLR $g$ prior $\times$ GEE  & 0.1508 & 0.1549 & 0.0889 & 0.0883 & 0.1607 & 0.1714 & 0.1018 & 0.1032 \\
 & BCLR $g$ prior $\times$ LR   & 0.1506 & 0.1548 & 0.0889 & 0.0885 & 0.1605 & 0.1714 & 0.1019 & 0.1030 \\
 & BCLR $g$ prior $\times$ GLMM & 0.1503 & 0.1550 & 0.0888 & 0.0884 & 0.1599 & 0.1714 & 0.1019 & 0.1032 \\
 & BCLR PMP $\times$ GEE        & 0.1025 & 0.1101 & 0.0821 & 0.0826 & 0.1046 & 0.1140 & 0.0928 & 0.0952 \\
 & BCLR PMP $\times$ LR         & 0.1025 & 0.1101 & 0.0820 & 0.0825 & 0.1045 & 0.1137 & 0.0929 & 0.0953 \\
 & BCLR PMP $\times$ GLMM       & 0.1026 & 0.1101 & 0.0820 & 0.0825 & 0.1045 & 0.1137 & 0.0929 & 0.0951 \\
 & BCLR Hybrid $\times$ GEE     & 0.1433 & 0.1524 & 0.0853 & 0.0867 & 0.1510 & 0.1657 & 0.0974 & 0.1012 \\
 & BCLR Hybrid $\times$ LR      & 0.1433 & 0.1522 & 0.0856 & 0.0868 & 0.1507 & 0.1659 & 0.0974 & 0.1012 \\
 & BCLR Hybrid $\times$ GLMM    & 0.1435 & 0.1521 & 0.0857 & 0.0869 & 0.1509 & 0.1661 & 0.0975 & 0.1009 \\
 & CLR                          & 0.1416 & 0.1575 & 0.0824 & 0.0885 & 0.1538 & 0.1768 & 0.0940 & 0.1015 \\
 & GEE                          & 0.0497 & 0.0566 & 0.0678 & 0.0684 & 0.0733 & 0.0740 & 0.0779 & 0.0783 \\
 & GLMM                         & 0.0938 & 0.0960 & 0.0753 & 0.0756 & 0.1005 & 0.1028 & 0.0850 & 0.0853 \\
 & LR                           & 0.0497 & 0.0567 & 0.0678 & 0.0684 & 0.0734 & 0.0741 & 0.0779 & 0.0784 \\ 
\midrule

\multirow{16}{*}{\textbf{500}}
 & BCLR Naive $\times$ GEE      & 0.0517 & 0.0510 & 0.0386 & 0.0392 & 0.0513 & 0.0533 & 0.0432 & 0.0445 \\
 & BCLR Naive $\times$ LR       & 0.0517 & 0.0509 & 0.0387 & 0.0392 & 0.0512 & 0.0534 & 0.0432 & 0.0446 \\
 & BCLR Naive $\times$ GLMM     & 0.0517 & 0.0509 & 0.0387 & 0.0392 & 0.0512 & 0.0534 & 0.0432 & 0.0445 \\
 & BCLR $g$ prior $\times$ GEE  & 0.0674 & 0.0658 & 0.0396 & 0.0405 & 0.0686 & 0.0705 & 0.0444 & 0.0461 \\
 & BCLR $g$ prior $\times$ LR   & 0.0673 & 0.0659 & 0.0396 & 0.0406 & 0.0686 & 0.0705 & 0.0444 & 0.0462 \\
 & BCLR $g$ prior $\times$ GLMM & 0.0675 & 0.0659 & 0.0397 & 0.0405 & 0.0685 & 0.0705 & 0.0445 & 0.0462 \\
 & BCLR PMP $\times$ GEE        & 0.0500 & 0.0506 & 0.0381 & 0.0390 & 0.0504 & 0.0531 & 0.0426 & 0.0446 \\
 & BCLR PMP $\times$ LR         & 0.0501 & 0.0505 & 0.0381 & 0.0391 & 0.0503 & 0.0531 & 0.0427 & 0.0446 \\
 & BCLR PMP $\times$ GLMM       & 0.0500 & 0.0506 & 0.0381 & 0.0391 & 0.0503 & 0.0531 & 0.0427 & 0.0446 \\
 & BCLR Hybrid $\times$ GEE     & 0.0659 & 0.0652 & 0.0389 & 0.0401 & 0.0664 & 0.0693 & 0.0435 & 0.0456 \\
 & BCLR Hybrid $\times$ LR      & 0.0659 & 0.0651 & 0.0390 & 0.0402 & 0.0666 & 0.0692 & 0.0436 & 0.0458 \\
 & BCLR Hybrid $\times$ GLMM    & 0.0659 & 0.0652 & 0.0390 & 0.0402 & 0.0663 & 0.0692 & 0.0435 & 0.0457 \\
 & CLR                          & 0.0633 & 0.0663 & 0.0386 & 0.0397 & 0.0646 & 0.0678 & 0.0442 & 0.0453 \\
 & GEE                          & 0.0244 & 0.0278 & 0.0333 & 0.0334 & 0.0480 & 0.0447 & 0.0396 & 0.0397 \\
 & GLMM                         & 0.0449 & 0.0457 & 0.0362 & 0.0363 & 0.0467 & 0.0472 & 0.0415 & 0.0415 \\
 & LR                           & 0.0245 & 0.0278 & 0.0333 & 0.0334 & 0.0480 & 0.0448 & 0.0396 & 0.0397 \\
\midrule
\end{tabular}
}
\caption{MSE (to four decimal places) for all settings averaged over the $N_{sim}$ iterations.}
\label{tab:mse}
\end{table*}

\begin{table*}[!ht]
\centering
\small
\resizebox{5in}{!}{%
\begin{tabular}{ll cccc cccc}
\multicolumn{2}{c}{} & \multicolumn{8}{c}{\textbf{Treatment effect}} \\ 
\cmidrule(lr){3-10}
\multicolumn{2}{c}{} & \multicolumn{4}{c}{$\beta_w = 0$} & \multicolumn{4}{c}{$\beta_w = 0.5$} \\ 
\cmidrule(lr){3-6} \cmidrule(lr){7-10}
\multicolumn{2}{c}{} & \multicolumn{8}{c}{\textbf{True function}} \\ 
\cmidrule(lr){3-10}
\multicolumn{2}{c}{} & \multicolumn{2}{c}{linear} & \multicolumn{2}{c}{non-linear} & \multicolumn{2}{c}{linear} & \multicolumn{2}{c}{non-linear} \\ 
\cmidrule(lr){3-4} \cmidrule(lr){5-6} \cmidrule(lr){7-8} \cmidrule(lr){9-10}
\textbf{N} & \textbf{Model} & \textbf{one} & \textbf{two} & \textbf{one} & \textbf{two} & \textbf{one} & \textbf{two} & \textbf{one} & \textbf{two} \\ 
\midrule

\multirow{16}{*}{\textbf{100}}
 & BCLR Naive $\times$ GEE    & 0.9550 & 0.9543 & 0.9530 & 0.9503 & 0.9550 & 0.9573 & 0.9481 & 0.9485 \\
 & BCLR Naive $\times$ LR     & 0.9615* & 0.9604* & 0.9485 & 0.9480 & 0.9635* & 0.9627* & 0.9477 & 0.9469 \\
 & BCLR Naive $\times$ GLMM   & 0.9578 & 0.9334* & 0.9457 & 0.9456 & 0.9598* & 0.9265* & 0.9500 & 0.9479 \\
 & BCLR $g$ prior $\times$ GEE & 0.9529 & 0.9540 & 0.9489 & 0.9501 & 0.9528 & 0.9551 & 0.9479 & 0.9483 \\
 & BCLR $g$ prior $\times$ LR  & 0.9691* & 0.9687* & 0.9466 & 0.9467 & 0.9716* & 0.9712* & 0.9526 & 0.9499 \\
 & BCLR $g$ prior $\times$ GLMM & 0.9685* & 0.9562 & 0.9442 & 0.9426 & 0.9724* & 0.9599* & 0.9531 & 0.9510 \\
 & BCLR PMP $\times$ GEE      & 0.9775* & 0.9787* & 0.9558 & 0.9574 & 0.9701* & 0.9747* & 0.9550 & 0.9542 \\
 & BCLR PMP $\times$ LR       & 0.9684* & 0.9689* & 0.9532 & 0.9567 & 0.9701* & 0.9703* & 0.9574 & 0.9584* \\
 & BCLR PMP $\times$ GLMM     & 0.9658* & 0.9489 & 0.9534 & 0.9520 & 0.9673* & 0.9456 & 0.9568 & 0.9586* \\
 & BCLR Hybrid $\times$ GEE   & 0.9681* & 0.9648* & 0.9506 & 0.9493 & 0.9590* & 0.9607* & 0.9535 & 0.9491 \\
 & BCLR Hybrid $\times$ LR    & 0.9765* & 0.9758* & 0.9526 & 0.9537 & 0.9773* & 0.9785* & 0.9576 & 0.9609* \\
 & BCLR Hybrid $\times$ GLMM  & 0.9766* & 0.9662* & 0.9514 & 0.9498 & 0.9776* & 0.9674* & 0.9588* & 0.9597* \\
 & CLR                       & 0.9839* & 0.9857* & 0.9582 & 0.9582 & 0.9823* & 0.9857* & 0.9644* & 0.9677* \\
 & GEE                       & 0.9447 & 0.9454 & 0.9464 & 0.9463 & 0.9195* & 0.9324* & 0.9418 & 0.9414* \\
 & GLMM                      & 0.9718* & 0.9705* & 0.9541 & 0.9536 & 0.9713* & 0.9700* & 0.9531 & 0.9520 \\
 & LR                        & 0.9805* & 0.9768* & 0.9558 & 0.9553 & 0.9703* & 0.9696* & 0.9510 & 0.9499 \\
\midrule

\multirow{16}{*}{\textbf{250}}
 & BCLR Naive $\times$ GEE    & 0.9504 & 0.9516 & 0.9511 & 0.9512 & 0.9414* & 0.9399* & 0.9480 & 0.9482 \\
 & BCLR Naive $\times$ LR     & 0.9599* & 0.9590* & 0.9501 & 0.9487 & 0.9535 & 0.9533 & 0.9478 & 0.9489 \\
 & BCLR Naive $\times$ GLMM   & 0.9511 & 0.9365* & 0.9477 & 0.9470 & 0.9488 & 0.9267* & 0.9464 & 0.9472 \\
 & BCLR $g$ prior $\times$ GEE & 0.9520 & 0.9523 & 0.9493 & 0.9486 & 0.9401* & 0.9402* & 0.9466 & 0.9463 \\
 & BCLR $g$ prior $\times$ LR  & 0.9517 & 0.9505 & 0.9467 & 0.9461 & 0.9496 & 0.9484 & 0.9461 & 0.9456 \\
 & BCLR $g$ prior $\times$ GLMM & 0.9486 & 0.9441 & 0.9465 & 0.9444 & 0.9462 & 0.9400* & 0.9481 & 0.9437 \\
 & BCLR PMP $\times$ GEE      & 0.9725* & 0.9736* & 0.9518 & 0.9513 & 0.9548 & 0.9551 & 0.9503 & 0.9490 \\
 & BCLR PMP $\times$ LR       & 0.9633* & 0.9627* & 0.9497 & 0.9520 & 0.9571 & 0.9565 & 0.9502 & 0.9495 \\
 & BCLR PMP $\times$ GLMM     & 0.9533 & 0.9434 & 0.9520 & 0.9498 & 0.9519 & 0.9362* & 0.9502 & 0.9479 \\
 & BCLR Hybrid $\times$ GEE   & 0.9449 & 0.9450 & 0.9502 & 0.9467 & 0.9357* & 0.9430 & 0.9458 & 0.9403* \\
 & BCLR Hybrid $\times$ LR    & 0.9530 & 0.9536 & 0.9481 & 0.9496 & 0.9519 & 0.9507 & 0.9489 & 0.9481 \\
 & BCLR Hybrid $\times$ GLMM  & 0.9521 & 0.9470 & 0.9480 & 0.9464 & 0.9506 & 0.9431 & 0.9482 & 0.9451 \\
 & CLR                       & 0.9569 & 0.9553 & 0.9522 & 0.9506 & 0.9553 & 0.9545 & 0.9511 & 0.9506 \\
 & GEE                       & 0.9503 & 0.9509 & 0.9497 & 0.9495 & 0.8832* & 0.9069* & 0.9431 & 0.9436 \\
 & GLMM                      & 0.9631* & 0.9625* & 0.9536 & 0.9535 & 0.9562 & 0.9565 & 0.9508 & 0.9508 \\
 & LR                        & 0.9816* & 0.9781* & 0.9560 & 0.9557 & 0.9465 & 0.9523 & 0.9499 & 0.9501 \\
\midrule

\multirow{16}{*}{\textbf{500}}
 & BCLR Naive $\times$ GEE    & 0.9488 & 0.9475 & 0.9486 & 0.9477 & 0.9434 & 0.9435 & 0.9430 & 0.9433 \\
 & BCLR Naive $\times$ LR     & 0.9574 & 0.9572 & 0.9473 & 0.9472 & 0.9567 & 0.9579 & 0.9455 & 0.9454 \\
 & BCLR Naive $\times$ GLMM   & 0.9510 & 0.9387* & 0.9447 & 0.9466 & 0.9545 & 0.9367* & 0.9443 & 0.9452 \\
 & BCLR $g$ prior $\times$ GEE & 0.9455 & 0.9443 & 0.9458 & 0.9463 & 0.9409* & 0.9407* & 0.9431 & 0.9443 \\
 & BCLR $g$ prior $\times$ LR  & 0.9479 & 0.9483 & 0.9459 & 0.9466 & 0.9534 & 0.9523 & 0.9462 & 0.9443 \\
 & BCLR $g$ prior $\times$ GLMM & 0.9460 & 0.9442 & 0.9460 & 0.9433 & 0.9516 & 0.9469 & 0.9465 & 0.9435 \\
 & BCLR PMP $\times$ GEE      & 0.9696* & 0.9703* & 0.9491 & 0.9491 & 0.9470 & 0.9462 & 0.9451 & 0.9454 \\
 & BCLR PMP $\times$ LR       & 0.9614* & 0.9600* & 0.9480 & 0.9486 & 0.9562 & 0.9551 & 0.9473 & 0.9473 \\
 & BCLR PMP $\times$ GLMM     & 0.9504 & 0.9431 & 0.9472 & 0.9474 & 0.9551 & 0.9430 & 0.9470 & 0.9467 \\
 & BCLR Hybrid $\times$ GEE   & 0.9501 & 0.9491 & 0.9488 & 0.9449 & 0.9489 & 0.9496 & 0.9437 & 0.9409* \\
 & BCLR Hybrid $\times$ LR    & 0.9503 & 0.9493 & 0.9472 & 0.9475 & 0.9542 & 0.9524 & 0.9468 & 0.9465 \\
 & BCLR Hybrid $\times$ GLMM  & 0.9470 & 0.9440 & 0.9468 & 0.9456 & 0.9515 & 0.9503 & 0.9479 & 0.9452 \\
 & CLR                       & 0.9510 & 0.9495 & 0.9496 & 0.9485 & 0.9556 & 0.9540 & 0.9474 & 0.9477 \\
 & GEE                       & 0.9460 & 0.9470 & 0.9489 & 0.9490 & 0.8229* & 0.8682* & 0.9390* & 0.9396* \\
 & GLMM                      & 0.9590* & 0.9582* & 0.9528 & 0.9526 & 0.9568 & 0.9564 & 0.9468 & 0.9470 \\
 & LR                        & 0.9813* & 0.9767* & 0.9558 & 0.9557 & 0.9091* & 0.9277* & 0.9450 & 0.9450 \\
\midrule
\end{tabular}
}
\caption{Coverage (to four decimal places) for all settings averaged over the $N_{sim}$ iterations. Asterisks indicate a rejection of the null that the coverage equals 95\% using a binomial exact test at the Bonferroni-corrected family-wise error rate of 5\%.}
\label{tab:coverage}
\end{table*}

\end{appendices}

\end{document}